\documentclass{pasj00}
\draft

\begin{document}
\SetRunningHead{Sato et al.}{Planetary Companions around Three
G and K Giants}
\Received{2007/10/02}
\Accepted{2008/01/23}

\title{Planetary Companions around Three Intermediate-Mass G and K Giants:
18 Del, $\xi$ Aql, and HD 81688}



%
 \author{
   Bun'ei \textsc{Sato},\altaffilmark{1}
   Hideyuki \textsc{Izumiura},\altaffilmark{2,3}
   Eri \textsc{Toyota},\altaffilmark{4}
   Eiji \textsc{Kambe},\altaffilmark{2}
   Masahiro \textsc{Ikoma},\altaffilmark{5}\\
   Masashi \textsc{Omiya},\altaffilmark{6}
   Seiji \textsc{Masuda},\altaffilmark{7}
   Yoichi \textsc{Takeda},\altaffilmark{8}
   Daisuke \textsc{Murata},\altaffilmark{4}\\
   Yoichi \textsc{Itoh},\altaffilmark{4}
   Hiroyasu \textsc{Ando},\altaffilmark{8}
   Michitoshi \textsc{Yoshida},\altaffilmark{2}
   Eiichiro \textsc{Kokubo},\altaffilmark{8}
   and 
   Shigeru \textsc{Ida}\altaffilmark{5}}
 \altaffiltext{1}{Global Edge Institute, Tokyo Institute of Technology,
2-12-1-S6-6 Ookayama, Meguro-ku, Tokyo 152-8550, Japan}
 \email{sato.b.aa@m.titech.ac.jp}
 \altaffiltext{2}{Okayama Astrophysical Observatory, National
Astronomical Observatory, Kamogata,
Okayama 719-0232, Japan}
 \email{izumiura@oao.nao.ac.jp,
        kambe@oao.nao.ac.jp, yoshida@oao.nao.ac.jp}
 \altaffiltext{3}{The Graduate University for Advanced Studies,
Shonan Village, Hayama, Kanagawa 240-0193, Japan}
 \altaffiltext{4}{Graduate School of Science, Kobe University,
1-1 Rokkodai, Nada, Kobe 657-8501, Japan}
 \email{toyota@kobe-u.ac.jp, yitoh@kobe-u.ac.jp,
daisuke@harbor.scitec.kobe-u.ac.jp}
\altaffiltext{5}{Department of Earth and Planetary Sciences,
Tokyo Institute of Technology, 2-12-1 Ookayama, Meguro-ku,
Tokyo 152-8551, Japan}
 \email{mikoma@geo.titech.ac.jp,ida@geo.titech.ac.jp}
\altaffiltext{6}{Department of Physics, Tokai University,
1117 Kitakaname, Hiratsuka, Kanagawa 259-1292, Japan}
 \email{ohmiya@peacock.rh.u-tokai.ac.jp}
 \altaffiltext{7}{Tokushima Science Museum, Asutamu Land Tokushima,
45-22 Kibigadani, Nato, Itano-cho, Itano-gun,
Tokushima 779-0111, Japan}
 \email{masuda@asutamu.jp}
 \altaffiltext{8}{National Astronomical Observatory, 2-21-1 Osawa,
Mitaka, Tokyo 181-8588, Japan}
 \email{takedayi@cc.nao.ac.jp, ando@optik.mtk.nao.ac.jp,
        kokubo@nao.ac.jp}


\maketitle

\begin{abstract}
We report the detection of 3 new extrasolar planets from the precise
Doppler survey of G and K giants at Okayama Astrophysical Observatory.
The host stars, namely, 18 Del (G6 III), $\xi$ Aql (K0 III) and
HD 81688 (K0 III--IV), are located at the clump region on the HR
diagram with estimated masses of $2.1-2.3M_{\odot}$. 18 Del b has
a minimum mass of $10.3 M_{\rm J}$ and resides in a nearly circular
orbit with period of 993 days, which is the longest one ever
discovered around evolved stars. $\xi$ Aql b and HD 81688 b have
minimum masses of 2.8 and 2.7 $M_{\rm J}$, and reside in nearly
circular orbits with periods of 137 and 184 days, respectively,
which are the shortest ones among planets around evolved stars.
All of the substellar companions ever discovered around possible
intermediate-mass ($1.7-3.9M_{\odot}$) clump giants have semimajor
axes larger than 0.68 AU, suggesting the lack of short-period planets.
Our numerical calculations suggest that Jupiter-mass planets within
about 0.5 AU (even up to 1 AU depending on the metallicity and
adopted models) around 2--3$M_{\odot}$ stars could be engulfed by
the central stars at the tip of RGB due to tidal torque from the
central stars. Assuming that most of the clump giants are post-RGB
stars, we can not distinguish whether the lack of short-period
planets is primordial or due to engulfment by central stars.
Deriving reliable mass and evolutionary status for evolved stars
is highly required for further investigation of formation and
evolution of planetary systems around intermediate-mass stars.
\end{abstract}

\section{Introduction}
Ongoing Doppler planet searches have discovered more than 200
extrasolar planets with various characteristics.\footnote
{See, e.g., tables at http://www.ciw.edu/boss/planets.html;
http://exoplanet.eu/}
Most of them are quite different from those in our solar-system:
the planets have minimum masses of 5 $M_{\oplus}$--15 $M_{\rm J}$
and are distributed in the range of orbital radii from 0.02 to 6 AU
with orbital eccentricities of 0--0.9 (e.g. Butler et al. 2006).
Distribution and correlation between these parameters
are now used to calibrate planet formation theories
by comparing with predictions from numerical simulations
taking account of key processes of planet formation
such as migration, disk lifetime, and variation of disk mass
(Ida \& Lin 2004; Alibert et al. 2005).

Properties of host stars also play an important role
to constrain planet formation mechanisms. For example, frequency
of planets is well correlated with metallicity of the host stars:
metal-rich stars tend to harbor more planets than metal-poor ones
do (Fischer \& Valenti 2005; Santos et al. 2005),
which supports the core-accretion scenario for the
mechanism of giant planet formation.
The host stars' mass can be another essential parameter.
Recently, Johnson et al. (2007a) showed that frequency of planets
around stars with $1.3\le M/M_{\odot}<1.9$ is as high as about
9\%, compared to about 4\% for solar-mass ($0.7\le M/M_{\odot}<1.3$)
stars and about 2\% for low-mass ($<0.7M_{\odot}$) K--M stars.
It suggests that giant planets are more abundant in more massive
stars probably because of their larger surface density of dust
in proto-planetary disks (Ida \& Lin 2005; Laughlin et al. 2004).
However, in higher-mass stars such as B--A dwarfs ($\ge2M_{\odot}$),
the shorter lifetime of the disk and the paucity of solid materials
close to the central star could reduce the abundance of giant
planets as a whole. The frequency of planets around
such massive stars has not been established yet.

Doppler planet searches around intermediate-mass ($\ge1.6M_{\odot}$)
stars have gradually expanded during these five years. Since massive
stars on the main-sequence (early-type stars) are unsuitable for
precise radial velocity measurements due to few absorption lines in
their spectra, which are often rotationally broadened,
major teams have targeted cool and slowly-rotating G and K giants
and subgiants, that is, massive stars in evolved stages
(Setiawan et al. 2005; Sato et al. 2003, 2007; Hatzes et al. 2005;
Johnson et al. 2007b; Lovis \& Mayor 2007; Niedzielski et al. 2007;
Liu et al. 2007).
They have succeeded in discovering 12 substellar companions so far
around stars with masses of 1.6--3.9$M_{\odot}$ including clump
giants and subgiants, and about a half of the host stars have
masses greater than 2$M_{\odot}$.
Although the number of planets is still small, the planets begin to
show different properties from those around low-mass stars:
high frequency of massive planets (Lovis \& Mayor 2007), lack of
inner planets (Johnson et al. 2007b), and low metallicity of
host stars (Pasquini et al. 2007).
These properties must reflect the history of formation and
evolution of planetary systems,
which are not necessarily the same as those for solar-type stars.
Planets around low-mass ($<1.6M_{\odot}$) giants have been also
discovered from precise radial velocity surveys
(Frink et al. 2002; Setiawan et al. 2003; Setiawan 2003;
Hatzes et al. 2003; D$\ddot{\rm{o}}$llinger et al. 2007).
Comparing orbital distribution of
the planets with those around low-mass dwarfs can give insight
into understanding of evolution of planetary systems.

In this paper, we report on the detection of 3 new extrasolar
planets around intermediate-mass G and K giants (18 Del,
$\xi$ Aql, HD 81688) from the Okayama
Planet Search Program (Sato et al. 2005).
We also update orbital parameters of
HD 104985 b, the first planet discovered around G giants
from our survey (Sato et al. 2003),
by using the data collected during the past six years.
Based on the extended sample, we discuss orbital properties of
planets around evolved stars taking account of evolution of
central stars.

\section{Observations}

Since 2001, we have been conducting a precise Doppler survey of
about 300 G and K giants (Sato et al. 2005) using a 1.88 m telescope,
the HIgh Dispersion Echelle Spectrograph (HIDES; Izumiura 1999),
and an iodine absorption cell (I$_2$ cell; Kambe et al. 2002)
at Okayama Astrophysical Observatory (OAO).
For precise radial velocity measurements, we set a wavelength range
to 5000--6100${\rm \AA}$, in which many deep and sharp I$_2$
lines exist, and a slit width to 200 $\mu$m ($0.76^{\prime\prime}$)
giving a spectral resolution ($R=\lambda/\Delta\lambda$) of 67000
by about 3.3 pixels sampling. We can typically obtain a
signal-to-noise ratio S/N$\gtrsim$200 pix$^{-1}$ for a $V\le 6$
star with an exposure time shorter than 30 min.
We have achieved a Doppler precision of about 6 m s$^{-1}$ over
a time span of 6 years using our own analysis software for modeling
an I$_2$-superposed stellar spectrum (Sato et al. 2002, 2005).

For abundance analysis, we take a pure (I$_2$-free) stellar
spectrum with the same wavelength range and spectral resolution as
those for radial velocity measurements. We also take a spectrum
covering Ca II H K lines in order to check the
chromospheric activity (not simultaneously obtained with the radial
velocity data) for stars showing large radial velocity variations.
In this case, we set the wavelength range to 3800--4500 ${\rm \AA}$
and the slit width to 250 $\mu$m giving a wavelength resolution of
50000. We can typically obtain S/N$\simeq$20 pix$^{-1}$ at
the Ca II H K line cores for a $B=6$ star with a 30 min exposure.
The reduction of echelle data (i.e. bias subtraction, flat-fielding,
scattered-light subtraction, and spectrum extraction) is performed
using the IRAF\footnote{IRAF is distributed by the National
Optical Astronomy Observatories, which is operated by the
Association of Universities for Research in Astronomy, Inc. under
cooperative agreement with the National Science Foundation,
USA.} software package in the standard way.

\section{Stellar Properties, Radial Velocities, and
Orbital Solutions}

\subsection{18 Del}

\begin{figure}
  \begin{center}
    \FigureFile(130mm,80mm){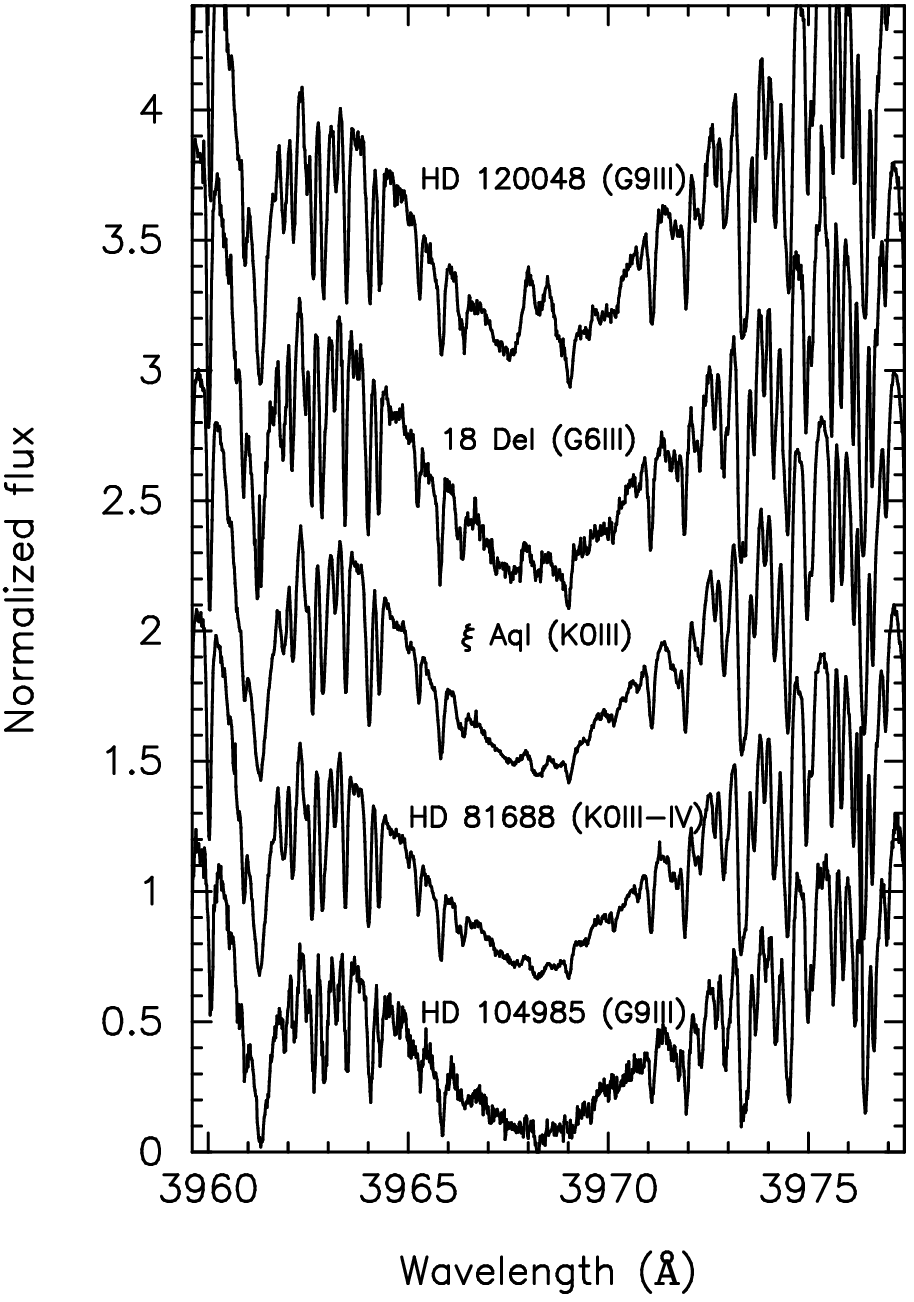}
  \end{center}
\caption{Spectra in the region of Ca H lines.
$\xi$ Aql, HD 81688, and HD 104985 show no significant emissions
in line cores. 18 Del shows slight core reversal but it is
not significant compared to that in HD 120048, a chromospheric
active star in our sample, which exhibits velocity scatter of
about 30 m s$^{-1}$ at most.
A vertical offset of about 0.7 is added to each spectrum.}\label{fig-CaH}
\end{figure}

18 Del (HR 8030, HD 199665, HIP 103527) is listed in the Hipparcos
catalog (ESA 1997) as a G6 III: giant star with
a $V$ magnitude $V=5.51$, a color index $B-V=0.934$,
and the Hipparcos parallax $\pi=13.68\pm0.70$ mas, corresponding
to a distance of 73.1$\pm$3.7 pc and an absolute magnitude $M_V=1.15$
taking account of correction of interstellar extinction $A_V=0.04$
based on the Arenou et al's (1992) table.
{\it Hipparcos} made a total of 175 observations of the star,
revealing a photometric stability down to $\sigma=0.007$ mag.
Figure \ref{fig-CaH} shows a Ca {\small II} H line for the star
obtained with HIDES revealing slight core reversal in the line.
X-ray luminosity for the star was derived to
$L_X=3.3\times10^{29}$ erg s$^{-1}$ from the $ROSAT$ measurements
(H$\ddot{\rm u}$nsch et al. 1998), suggesting that the
star is slightly chromospherically active. However, the reversal
is not significant compared to those in other chromospherically
active stars in our sample such as HD 120048 (Figure \ref{fig-CaH}),
which shows velocity scatter of about 30 m s$^{-1}$.
Thus, although the correlation between
chromospheric activity and intrinsic radial velocity ``jitter'' for
giants have not been well established yet, the jitter of 18 Del
is probably expected to be no larger than that of HD 120048.
Further discussions are presented in Section 4.

\begin{figure}
  \begin{center}
    \FigureFile(120mm,80mm){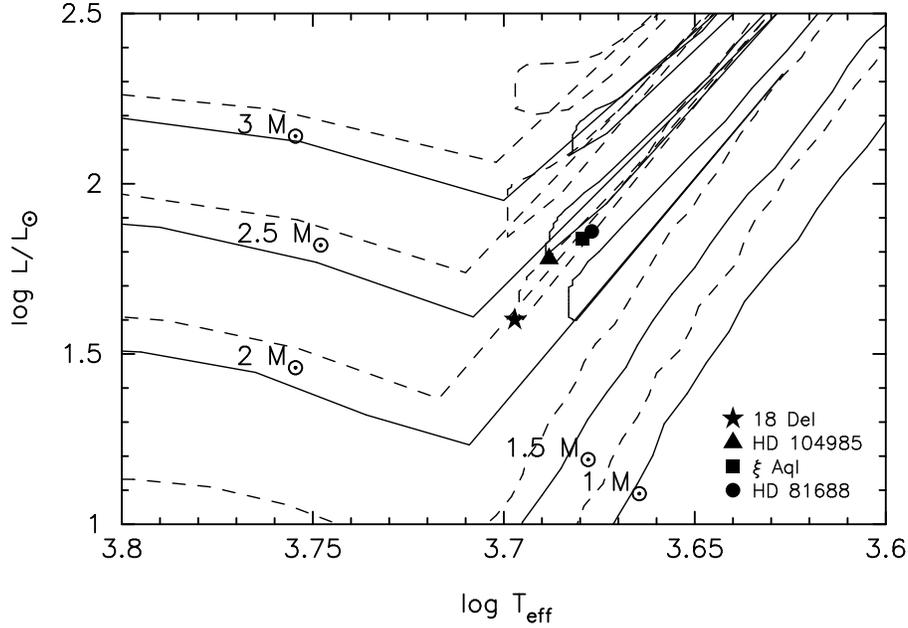}
  \end{center}
\caption{HR diagram of the planet-harboring stars presented in this paper.
Evolutionary tracks from Lejeune and Schaerer (2001)
for stars with $Z=0.02$ (solar metallicity; solid
lines) and $Z=0.008$ (dashed lines) of masses between 1 and 3
$M_{\odot}$ are also shown.}\label{fig-HRD}
\end{figure}

The atmospheric parameters and the Fe abundance of the star were
determined based on the spectroscopic approach using the equivalent 
widths of well-behaved Fe I and Fe II lines (cf. Takeda et al. 
2002 for a detailed description of this method).
We obtained 
$T_{\rm eff}$ = 4979 K,
$\log g$ = 2.82 cm~s$^{-2}$, 
$v_{\rm t}$ = 1.22 km~s$^{-1}$, 
and [Fe/H] = $-$0.05 for the star.
The bolometric correction was estimated
to $B.C.=-0.39$ based on the Kurucz (1993)'s theoretical calculation.
With use of these parameters and theoretical evolutionary tracks
of Lejeune \& Schaerer (2001), we derived the fundamental stellar
parameters, $L=40L_{\odot}$, $R=8.5R_{\odot}$, and
$M=2.3M_{\odot}$, as summarized in Table \ref{tbl-stars}.
The procedure described here is the same as that adopted in
Takeda et al. (2005) (see subsection 3.2 of Takeda et al. (2005)
and Note of Table \ref{tbl-stars} for uncertainties involved
in the stellar parameters).
The position of the star on the HR diagram is shown in Figure
\ref{fig-HRD} together with other planet-harboring stars in this
paper.
da Silva et al. (2006) obtained 2.13 $\pm$ 0.13 $M_{\odot}$
for the mass of the star based on $T_{\rm eff}$ = 5089 $\pm$ 70 K,
$\log g$ = 2.93 $\pm$ 0.08 cm~s$^{-2}$, and [Fe/H] = 0.05$\pm$0.05.
Although the $T_{\rm eff}$ is about 100 K higher than that we obtained,
the mass for the star reasonably agrees with our estimate.

\begin{figure}
  \begin{center}
    \FigureFile(120mm,80mm){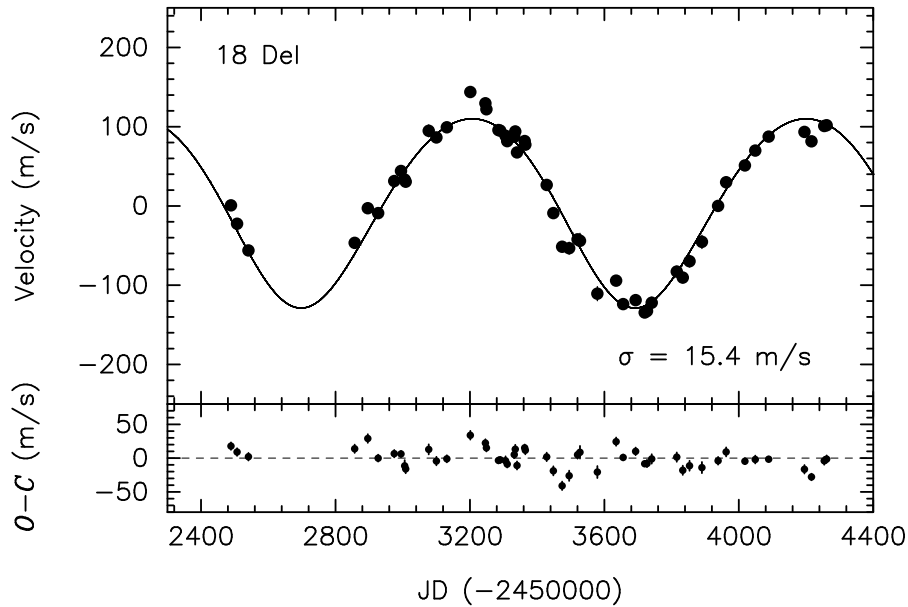}
  \end{center}
\caption{{\it Top}: Observed radial velocities of 18 Del (dots). 
The Keplerian orbital fit is shown by the solid line. The period
is 993 days, the velocity semiamplitude is 119 m s$^{-1}$,
and the eccentricity is 0.08. Adopting a stellar mass of 2.3 $M_{\odot}$,
we obtain the minimum mass for the companion of 10.3 $M_{\rm J}$,
and semimajor axis of 2.6 AU. {\it Bottom}: 
Residuals to the Keplerian fit. The rms to the fit is 15.4 m s$^{-1}$.}
\label{fig-HD199665}
\end{figure}

We collected a total of 51 radial velocity data of 18 Del
between 2002 August and 2007 June, with a typical S/N of 200 pix$^{-1}$
for an exposure time of about 900 s.
The observed radial velocities are shown in Figure \ref{fig-HD199665}
and are listed in Table \ref{tbl-HD199665} together with their
estimated uncertainties, which were derived from an ensemble of
velocities from each of $\sim$200 spectral regions
(each 4--5${\rm \AA}$ long) in every exposure.
The sinusoidal variability in the radial velocities is visible to the
eye and it can be well
fitted by a Keplerian orbit with a period $P=993.3\pm3.2$ days,
a velocity semiamplitude $K_1=119.4\pm1.3$ m s$^{-1}$, and an
eccentricity $e=0.08\pm0.01$.
The resulting model is shown in Figure \ref{fig-HD199665} overplotted
on the velocities, and its parameters are listed in
Table \ref{tbl-planets}.
The uncertainty of each parameter was estimated using a Monte
Carlo approach. We generated 100 fake datasets
by adding random Gaussian noise corresponding to velocity measurement
errors to the observed radial velocities in each set, then found the
best-fit Keplerian parameters for each, and examined the distribution of
each of the parameters.
The rms scatter of the residuals to the Keplerian fit is
15.4 m s$^{-1}$, which is comparable to the scatters of
giants with the same $B-V$ as 18 Del in our sample
(Sato et al. 2005). Adopting a stellar mass of 2.3 $M_{\odot}$,
we obtain a minimum mass for the companion
$m_2\sin i=10.3$ $M_{\rm J}$ and a semimajor axis
$a=2.6$ AU. The companion has the longest orbital
period ever discovered around evolved stars.

\begin{figure}
  \begin{center}
    \FigureFile(120mm,80mm){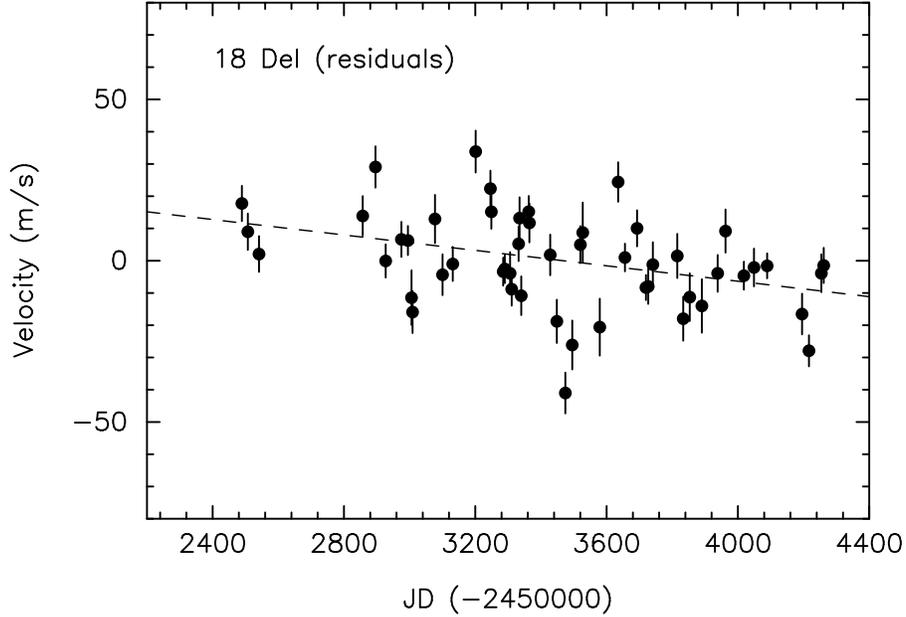}
  \end{center}
\caption{Linear fit to the residuals to the Keplearian orbit for
18 Del. The dashed line shows the best-fit linear trend
corresponding to $-4.4$ m s$^{-1}$ yr$^{-1}$.}
\label{fig-HD199665-res}
\end{figure}

We found that the residuals showed a decreasing trend
with a slope of about $-$4 m s$^{-1}$ yr$^{-1}$, suggesting
the existence of an outer companion
(Figure \ref{fig-HD199665-res}). Keplerian orbital
fit including a linear trend slightly improves the quality of
fit, decreasing the rms scatter from 15.4 m s$^{-1}$ to
13.6 m s$^{-1}$ and the reduced ${\chi^2}$ from
$\chi^2_{notrend}=6.3$ to $\chi^2_{trend}=5.0$.
To assess the significance of this trend, we evaluated a
false-alarm probability ($FAP$),
the probability that noise mimics the observed trend,
by using a bootstrap analysis which is the same
as adopted in Wright et al. (2007). We scrambled the residuals
in a random manner while keeping fixed the observation time,
and created a mock set of radial velocities by adding the residuals
back to the best-fit Keplerian radial velocity curve. We created
100 such mock data sets, and obtained
$\Delta\chi^2=\chi^2_{trend}-\chi^2_{notrend}$ for each set.
Eight of the 100 data sets showed $\Delta\chi^2$ less than
that for the original data set, which means the $FAP$ is 8\%.
Considering the $FAP$, we can not say that the observed trend
is significant at this stage.

\subsection{$\xi$ Aql}

$\xi$ Aql (HR 7595, HD 188310, HIP 97938) is a K0 III giant
star with a $V$ magnitude $V=4.71$, a color index $B-V=1.023$,
and a trigonometric parallax $\pi=15.96\pm1.01$ mas (ESA 1997),
placing the star at a distance of 62.7$\pm$4.0 pc. The distance and
an estimated interstellar extinction $A_V=0.10$ (Arenou et al. 1992)
yield an absolute magnitude for the star $M_V=0.63$.
{\it Hipparcos} made a total of 98 observations of the star, revealing
a photometric stability down to $\sigma=0.004$ mag.
Ca {\small II} H K lines of the star show no significant
emission in the line cores as shown in the Figure \ref{fig-CaH},
suggesting that the star is chromospherically inactive.
The atmospheric parameters, Fe abundance, and other fundamental
parameters of the star are listed in Table \ref{tbl-stars}.
The [Fe/H] of $-0.21$ is consistent with the
value of $-0.15$ from Taylor (1999) within the error.
Gray and Brown (2001) obtained $T_{\rm eff}=4670$ K for the star
based on line-depth-ratio analysis, which is $\sim100$ K lower than
our estimate (4780 K). Uncertainties in the mass and the radius
are considered to be similar to those for 18 Del
(see Note of Table \ref{tbl-stars}).

\begin{figure}
  \begin{center}
    \FigureFile(120mm,80mm){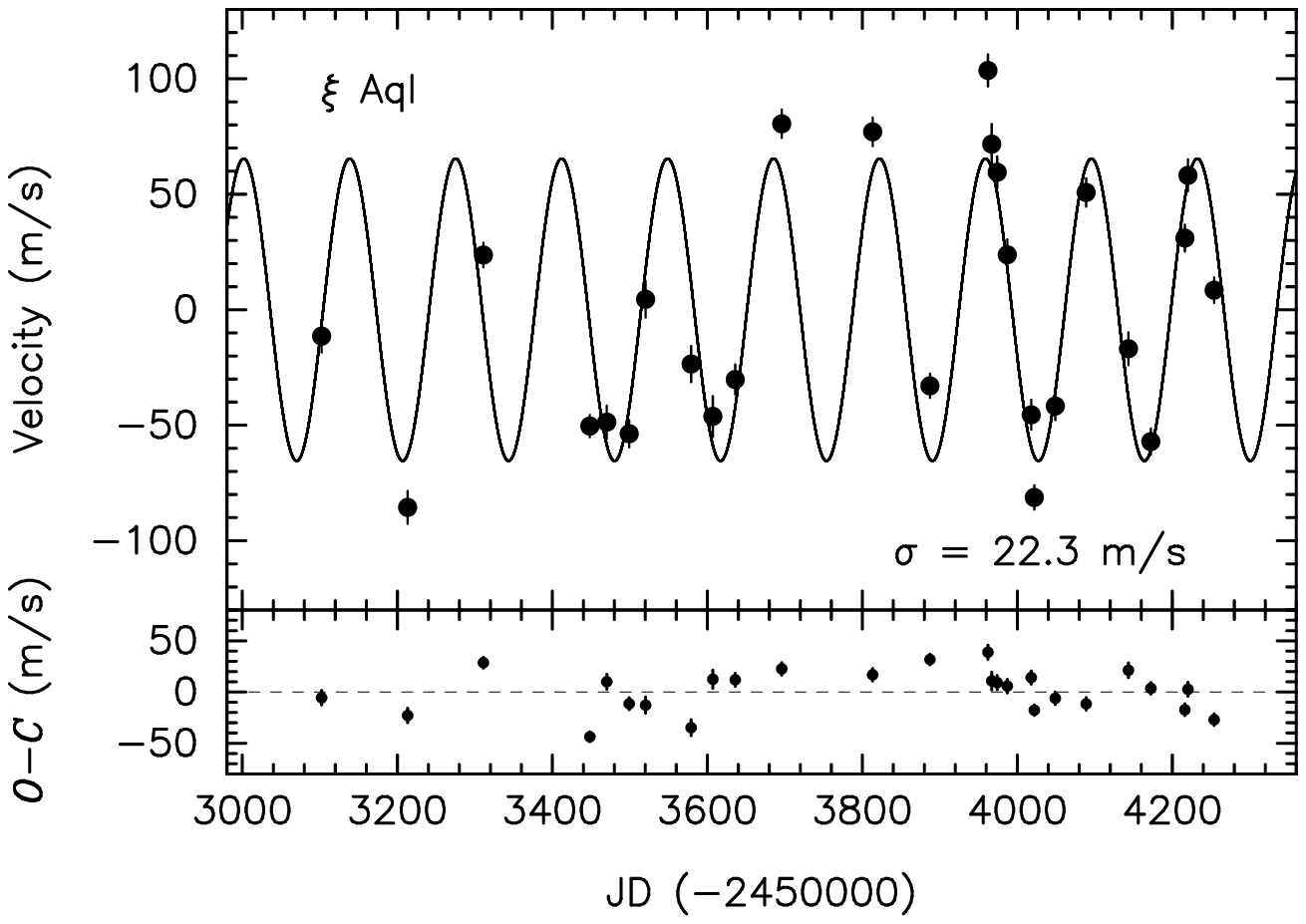}
  \end{center}
\caption{{\it Top}: Observed radial velocities of $\xi$ Aql (dots). 
The Keplerian orbital fit is shown by the solid line. The period
is 136.8 days and the velocity semiamplitude is 65 m s$^{-1}$
(the eccentricity is fixed to zero). Adopting a stellar mass of
2.2 $M_{\odot}$, we obtain the minimum mass for the companion
of 2.8 $M_{\rm J}$, and semimajor axis of 0.68 AU. {\it Bottom}: 
Residuals to the Keplerian fit. The rms to the fit is 22.3 m s$^{-1}$.}
\label{fig-HD188310}
\end{figure}

We collected a total of 26 radial velocity data of $\xi$ Aql
between 2004 April and 2007 June, with a typical S/N of 200
pix$^{-1}$ for an exposure time of about 300 s.
The observed radial velocities are shown in Figure \ref{fig-HD188310}
and are listed in Table \ref{tbl-HD188310} together with their
estimated uncertainties. Lomb-Scargle periodogram (Scargle 1982)
of the data exhibits a dominant peak at a period of 137 days.
To assess the significance of this periodicity, we estimated $FAP$,
using a bootstrap randomization method in which the observed radial
velocities were randomly redistributed,
keeping fixed the observation time. We generated 10$^5$ fake datasets
in this way, and applied the same periodogram analysis to them. Only 2
fake datasets exhibited a periodogram power higher than the observed
dataset. Therefore, the $FAP$ is $2\times10^{-5}$.
The observed radial velocities can be well fitted
by a circular orbit with a period $P=136.75\pm0.25$ days and
a velocity semiamplitude $K_1=65.4\pm1.7$ m s$^{-1}$.
The resulting model is shown in Figure \ref{fig-HD188310}, and its
parameters are listed in Table \ref{tbl-planets}. The uncertainty of
each parameter was estimated using a Monte Carlo approach as
described in Section 3.1.
The rms scatter of the residuals to the Keplerian fit was
22.3 m s$^{-1}$, which is slightly larger than the typical
scatter of giants with $B-V\simeq 1.0$ in our sample
(Sato et al. 2005). Adopting a stellar mass of 2.2 $M_{\odot}$,
we obtain a minimum mass for the companion of
$m_2\sin i=2.8M_{\rm J}$ and a semimajor axis of
$a=0.68$ AU. The companion has the shortest orbital
period ever discovered among evolved stars.

\subsection{HD 81688}

HD 81688 (HR 3743, HIP 46471) is classified in the Hipparcos
catalog (ESA 1997) as a K0 III--IV star with
a $V$ magnitude $V=5.40$, a color index $B-V=0.993$.
The Hipparcos parallax $\pi=11.33\pm0.84$ mas corresponds to a distance
of 88.3$\pm$6.5 pc and yields an absolute magnitude $M_V=0.57$
corrected by interstellar extinction $A_V=0.10$ (Arenou et al. 1992).
{\it Hipparcos} made a total of 107 observations of the star, revealing
a photometric stability down to $\sigma=0.006$ mag.
Ca {\small II} H K lines of the star show no significant
emission in the line cores, suggesting that the star is
chromospherically inactive (Figure \ref{fig-CaH}).
The atmospheric parameters, Fe abundance, and other fundamental
parameters of the star are listed in Table \ref{tbl-stars}.
Mishenina et al. (2006) obtained $T_{\rm eff}=4789$ K
(from line-depth-ratio analysis),
$\log g$ = 2.3 cm~s$^{-2}$, $v_{\rm t}$ = 1.3 km~s$^{-1}$, 
and [Fe/H] = $-$0.23 for the star. While the [Fe/H] is $\sim0.1$
dex higher than our estimate, other parameters reasonably
agree with those we obtained. Although uncertainties in
the mass and the radius are considered to be similar to
those for 18 Del, systematic error up to $\sim0.5M_{\odot}$
could exist depending on
adopted stellar evolutionary models (see Note of Table \ref{tbl-stars}).

\begin{figure}
  \begin{center}
    \FigureFile(120mm,80mm){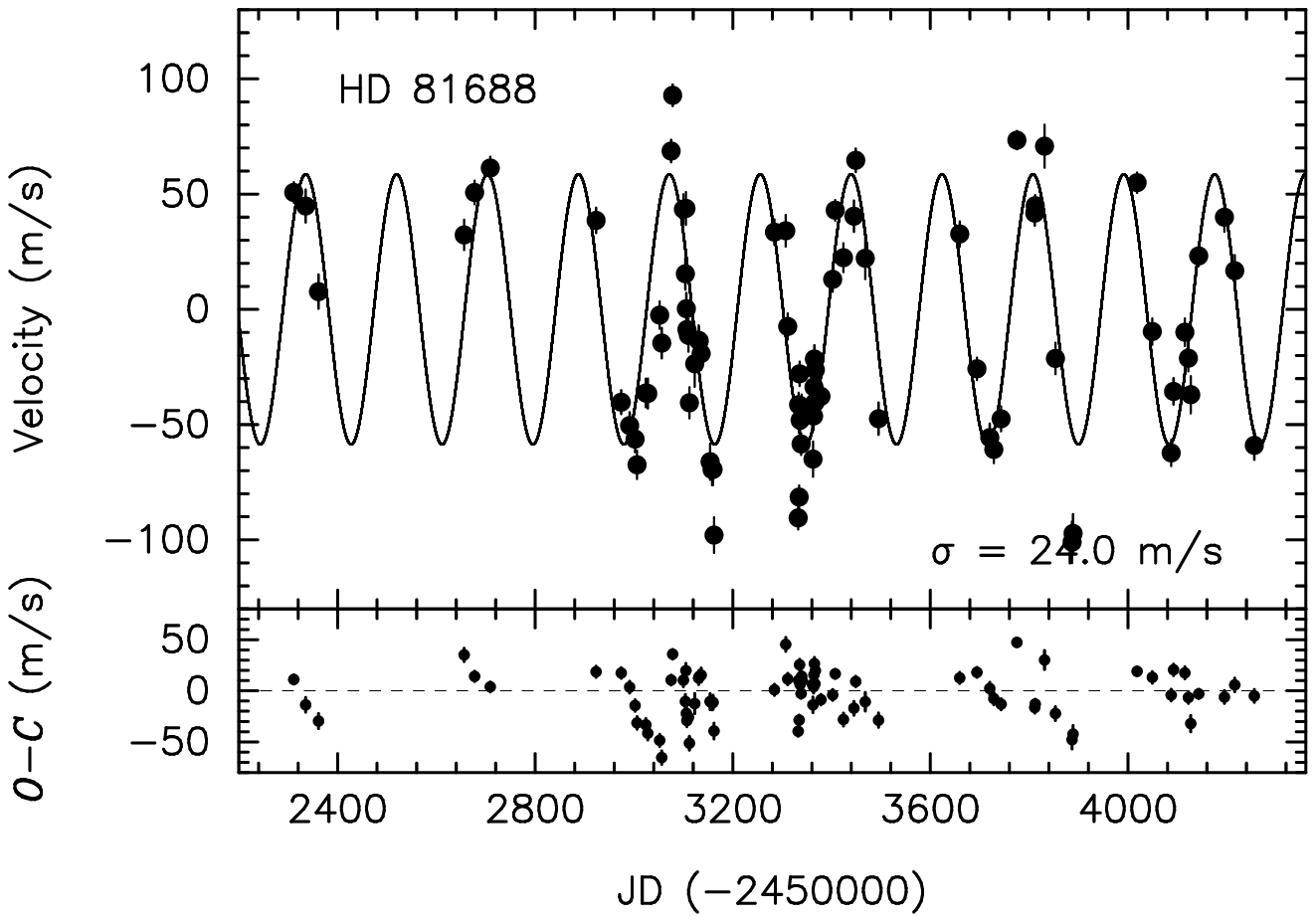}
  \end{center}
\caption{{\it Top}: Observed radial velocities of HD 81688 (dots). 
The Keplerian orbital fit is shown by the solid line. The period
is 184.0 days and the velocity semiamplitude is 59 m s$^{-1}$
(the eccentricity is fixed to zero). Adopting a stellar mass of
2.1 $M_{\odot}$, we obtain the minimum mass for the companion
of 2.7 $M_{\rm J}$, and semimajor axis of 0.81 AU. {\it Bottom}: 
Residuals to the Keplerian fit. The rms to the fit is 24.0 m s$^{-1}$.}
\label{fig-HD81688}
\end{figure}

We collected a total of 81 radial velocity data of HD 81688
between 2003 March and 2007 April, with a typical S/N
of 200 pix$^{-1}$ for an exposure time of about 900 s.
The observed radial velocities are shown in Figure \ref{fig-HD81688}
and are listed in Table \ref{tbl-HD81688} together with their
estimated uncertainties. Lomb-Scargle periodogram (Scargle 1982)
of the data exhibits a dominant peak at a period of 182 days
with a $FAP<1 \times 10^{-5}$, which was estimated by the same
method as described in Section 3.1.
The observed radial velocities can be well fitted by a circular
orbit with a period $P=184.02\pm0.18$ days and a velocity
semiamplitude $K_1=58.58\pm0.97$ m s$^{-1}$.
The resulting model is shown in Figure \ref{fig-HD81688}, and its
parameters are listed in Table \ref{tbl-planets}. The uncertainty
of each parameter was estimated using a Monte Carlo approach
as described in Section 3.1.
Adopting a stellar mass of 2.1 $M_{\odot}$,
we obtain a minimum mass for the companion of
$m_2\sin i=2.7$ $M_{\rm J}$ and a semimajor axis of
$a=0.81$ AU.

The residuals to the Keplerian fit exhibit non-random variations
in some periods of time, which may be due to stellar activity or
additional companions. We performed periodogram analysis
(Scargle 1982) to the residuals and found peaks at periods around
13, 25, and 55 days. However, the $FAP$'s for the peaks are larger
than 0.5, which is not considered to be significant at this stage
(some of them my be affected by aliasing). More dense sampling of
data will help discriminate if the periods are real or not.

\subsection{HD 104985}

HD 104985 (HR 4609, HIP 58952) is the first planet-harboring
star discovered from our survey (Sato et al. 2003).
It is classified in the Hipparcos
catalog (ESA 1997) as a G9 III giant star with
a $V$ magnitude $V=5.78$, a color index $B-V=1.029$,
and a parallax $\pi=9.80\pm0.52$ mas, corresponding to a distance
of 102.0$\pm$5.4 pc and an absolute magnitude $M_V=0.74$.

The atmospheric parameters of the star were updated
by Takeda et al. (2005) from those
listed in the discovery paper by Sato et al. (2003) to:
$T_{\rm eff}$ = 4877 K,
$\log g$ = 2.85 (cm~s$^{-2}$), 
$v_{\rm t}$ = 1.31 (km~s$^{-1}$), 
and [Fe/H] = $-0.15$.
Based on these parameters and the bolometric correction of
$B.C.=-0.43$ (Kurucz 1993),
Takeda et al. (2005) obtained the fundamental stellar parameters
of $L=60L_{\odot}$, $R=11R_{\odot}$, and $M=2.3M_{\odot}$.
\footnote{Sato et al. (2003) obtained a mass for the star
of 1.6 $M_{\odot}$ based on the metallicity [Fe/H]$=-0.35$ and
evolutionary tracks from Girardi et al. (2000). The tracks tend
to give $\lesssim0.5M_{\odot}$ lower mass compared to those
from Lejeune and Schaerer (2001) for $\lesssim2M_{\odot}$ giants with
$Z=0.008$ ([Fe/H]$=-0.4$).}

\begin{figure}
  \begin{center}
    \FigureFile(120mm,80mm){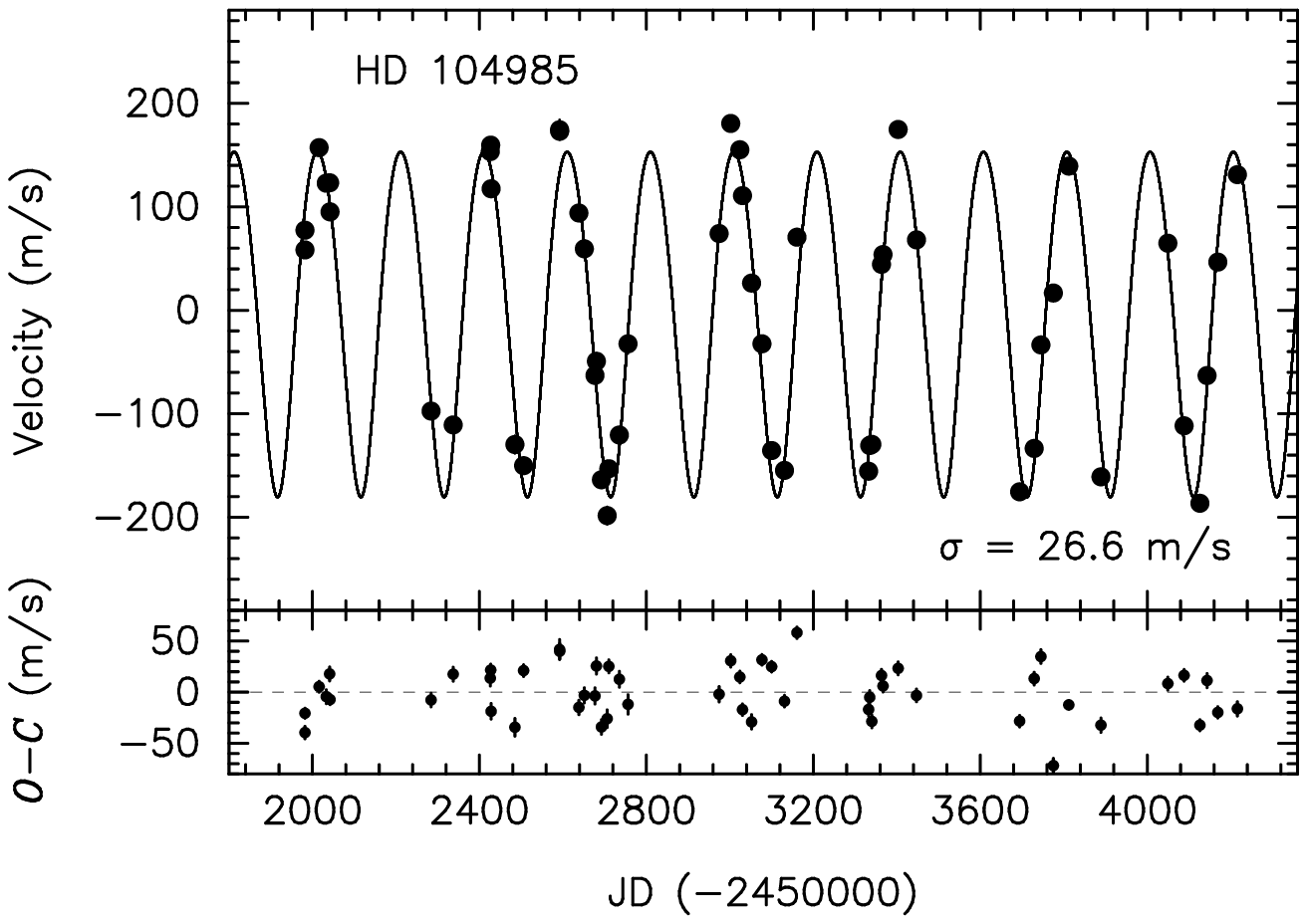}
  \end{center}
\caption{{\it Top}: Observed radial velocities of HD 104985 (dots). 
The Keplerian orbital fit is shown by the solid line. The period
is 199.51 days and the velocity semiamplitude is 167 m s$^{-1}$,
and the eccentricity is 0.09. Adopting a stellar mass of
2.3 $M_{\odot}$, we obtain the minimum mass for the companion
of 8.3 $M_{\rm J}$, and semimajor axis of 0.95 AU. {\it Bottom}: 
Residuals to the Keplerian fit. The rms to the fit is 26.6 m s$^{-1}$.}
\label{fig-HD104985}
\end{figure}

After the discovery of the planet around HD 104985 in 2003,
we have continued observations of the star and collected a
total of 52 data points between 2001 March and 2007 April.
The observed radial velocities are shown in Figure
\ref{fig-HD104985} and are listed in Table \ref{tbl-HD104985}
together with their estimated uncertainties.
Based on the extended data set, we updated the orbital parameters
of the planet: $P=199.505\pm0.085$ days,
$K_1=166.8\pm1.3$ m s$^{-1}$, and $e=0.090\pm0.009$.
The resulting Keplerian model is shown in Figure \ref{fig-HD81688},
and the parameters are listed in Table \ref{tbl-planets}.
Adopting a stellar mass of $2.3M_{\odot}$, 
we obtain a minimum mass for the companion of
$m_2\sin i=8.3M_{\rm J}$ and a semimajor axis of
$a=0.95$ AU. We can not find any additional periodic
signals or long-term trend in their radial velocities for now.

\section{Line Shape Analysis}

To investigate other causes producing apparent radial velocity
variations such as pulsation and rotational modulation rather
than orbital motion, spectral line shape analysis was performed
with use of high resolution stellar templates followed by the
technique of Sato et al. (2007). In our technique, we extract
a high resolution iodine-free stellar template from several stellar
spectra contaminated by iodine lines (Sato et al. 2002).
Basic procedure of the technique is as follows; first, we model observed
star+I$_2$ spectrum in a standard manner but using the initial
guess of the intrinsic stellar template spectrum.
Next we take the difference between the observed star+I$_2$
spectrum and the modeled one. Since the difference is mainly considered
to be due to an imperfection of the initial guess of the stellar
template spectrum, we revise the initial guess taking account of the
difference and model the observed star+I$_2$ spectrum using the revised
guess of the template. We repeat this process until we obtain sufficient
agreement between observed and modeled spectrum. We take average
of thus obtained stellar templates from several observed star+I$_2$
spectra to increase S/N ratio of the template. Details of this
technique are described in Sato et al. (2002).

For spectral line shape analysis, we extracted two stellar templates
from several
star+I$_2$ spectra at the peak and valley phases of observed radial
velocities for each star. Then, cross correlation
profiles of the two templates were calculated for 30--50 spectral
segments (4--5${\rm \AA}$ width each) in which severely blended
lines or broad lines were not included.
Three bisector quantities were calculated for the cross correlation
profile of each segment: the velocity span (BVS), which is the
velocity difference between two flux levels of the bisector;
the velocity curvature (BVC), which is the difference of the
velocity span of the upper half and lower half of the bisector;
and the velocity displacement (BVD), which is the average of
the bisector at three different flux levels.
We used flux levels of 25\%, 50\%, and 75\% of the cross
correlation profile to calculate the above quantities.
Resulting bisector quantities for 18 Del, $\xi$ Aql, HD 81688,
and HD 104985 are listed in Table \ref{tbl-bisector}.
As expected from the planetary hypothesis, both of the BVS and
the BVC are identical to zero, which means that the cross
correlation profiles are symmetric, and the average BVD is
consistent with the velocity difference between the two templates
at the peak and valley phases of observed radial velocities
($\simeq 2K_1$).
The BVS for HD 104985 is slightly large compared to those for
other stars, suggesting the higher intrinsic variability for
HD 104985. It may be consistent with the large rms scatters of
the residuals to the Keplerian fit ($\sigma=26.6$ m s$^{-1}$)
for the star. However, the BVS value is only one thirtieth
of the BVD and thus it is unlikely that the observed radial
velocity variations are produced by the intrinsic activity
such as pulsation or rotational modulation. Based on these
results, we conclude that the radial velocity variability
observed in these 4 stars are best explained by orbital motion.

\section{Discussion}

\begin{figure}
  \begin{center}
    \FigureFile(110mm,80mm){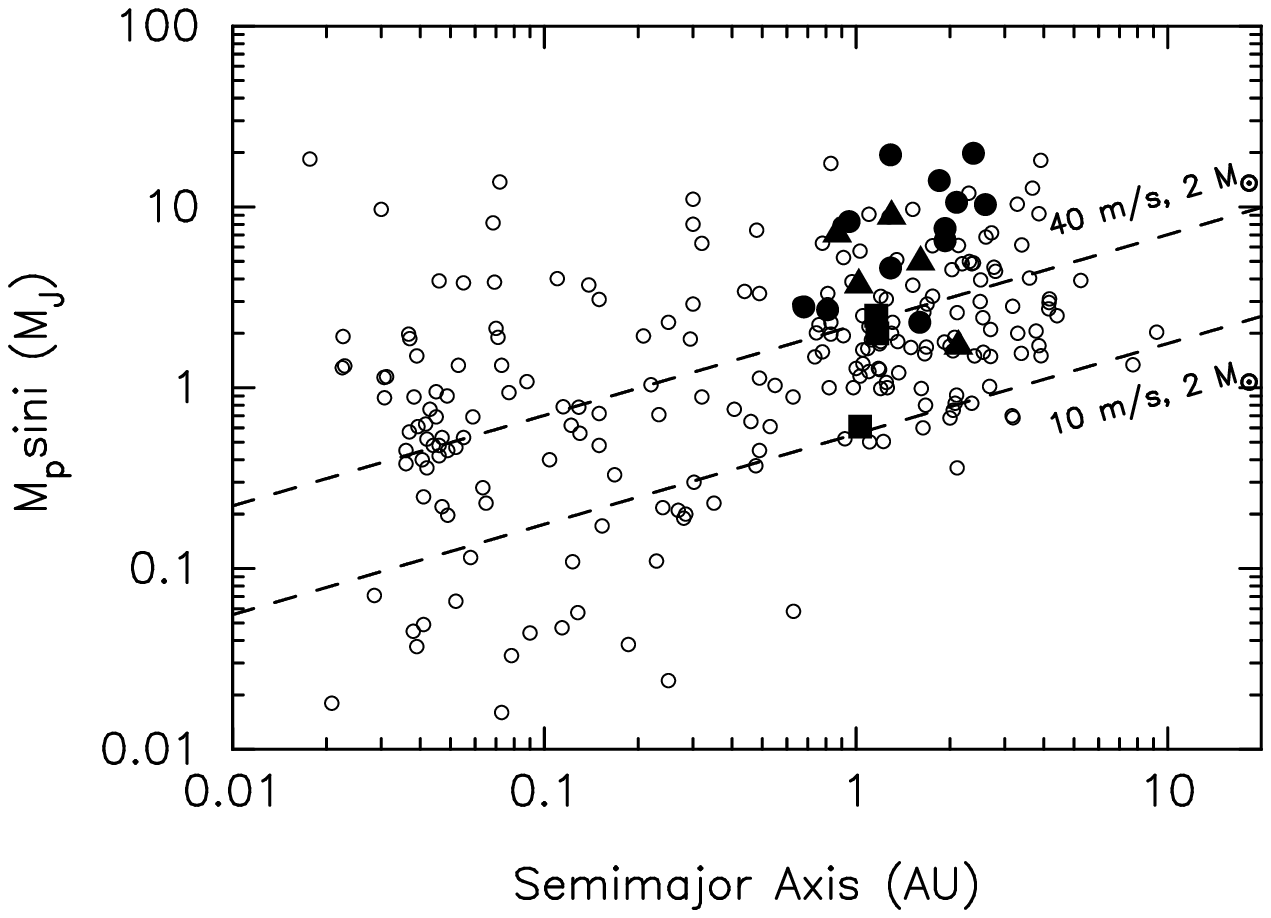}
  \end{center}
\caption{Mass of extrasolar planets plotted against
semimajor axis. Planets around low-mass ($<1.6M_{\odot}$)
giants, intermediate-mass (1.6--1.9$M_{\odot}$) subgiants,
and clump giants (1.7--3.9$M_{\odot}$), are plotted
by filled triangles, filled squares, and filled circles,
respectively.
Planets around solar-type dwarfs are plotted by open circles.}
\label{fig-avsm}
\end{figure}

\begin{figure}
  \begin{center}
    \FigureFile(110mm,80mm){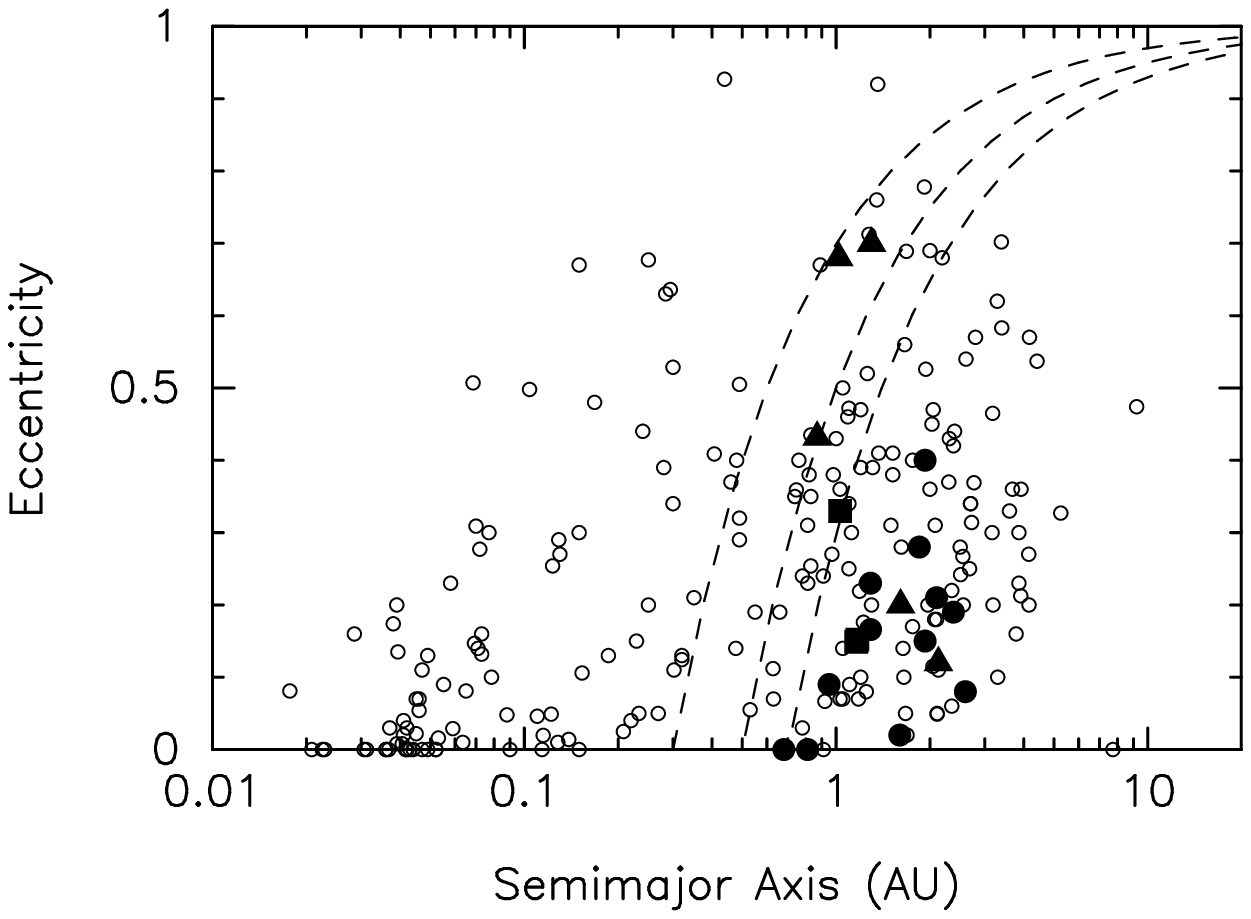}
  \end{center}
\caption{Eccentricity of extrasolar planets plotted against
semimajor axis. Planets around low-mass ($<1.6M_{\odot}$)
giants, intermediate-mass (1.6--1.9$M_{\odot}$) subgiants,
and clump giants (1.7--3.9$M_{\odot}$), are plotted
by filled triangles, filled squares, and filled circles,
respectively.
Planets around solar-type dwarfs are plotted by open circles.
Dashed lines express the periastron distance ($q=a(1-e)$)
of 0.3, 0.5, 0.7 AU, respectively, from the left.}
\label{fig-avse}
\end{figure}

We discovered a total of 6 substellar companions around
G and K giants so far from our Okayama Planet Search Program
(Sato et al. 2003, 2007; Liu et al. 2007; this work).
The host stars are located at the clump region on the HR
diagram and their masses are estimated to 2.1--2.7
$M_{\odot}$. When we include 6 more planets discovered
around possible clump giants by other teams
(Table \ref{tbl-pllist}), the mass of the host stars ranges
from 1.7 to 3.9 $M_{\odot}$. These discoveries definitely
indicate that planets can form around intermediate-mass
stars, such as B--A dwarfs, as well as around low-mass ones.
The extended sample enables us to clarify the properties of
the planets around intermediate-mass clump giants; the planets
have minimum masses of 2.3--19.8$M_{\rm J}$, semimajor axes
of 0.68--2.6 AU, and eccentricities of 0--0.4.
In Figure \ref{fig-avsm}, we plotted the minimum mass against
the semimajor axis.
Since intrinsic variability in radial velocity of clump giants
is typically 10--20 m s$^{-1}$ (Sato et al. 2005), it is normally
difficult to detect planets with $\lesssim 2M_{\rm J}$
at $\simeq$1 AU around a 2$M_{\odot}$ star, which can produce
radial velocity semiamplitude of 40 m s$^{-1}$ at most.
Such lower-mass planets can be detected around stars with small
intrinsic variability ($<$10 m s$^{-1}$) like subgiants
(Johnson et al. 2007b).
While the largest semimajor axis of 2.6 AU is limited by the time
baseline of the current surveys, the lack of short-period planets
with $a\lesssim$0.7 AU appears to be real at least for relatively
massive planets because stellar radius of clump
giants is typically 10--20$R_{\odot}$, which correspond to
0.05--0.1 AU, and thus we should be able to find planets with
0.1$\lesssim a\lesssim$0.7 AU if they exist.
In Figure \ref{fig-avse}, we plotted the eccentricity against
the semimajor axis together with lines expressing different
periastron distances ($q=a(1-e)$). From the view point of orbital
evolution of planets, periastron distance is more essential
rather than semimajor axis because tidal interaction between
a planet and a central star strongly depends on distance between
them. As shown in the figure, all the companions
around clump giants have $q\ge0.68$ AU,
while those around intermediate-mass subgiants
(1.6--1.9$M_{\odot}$) and low-mass K giants
($<1.6M_{\odot}$) have $q\ge0.69$ AU and $\ge0.33$ AU,
respectively.
Since a lot of planets with $q\le$0.3 AU have been found
around solar-type dwarfs (open circles), the lack of inner planets
around low-mass K giants can be due to engulfment by the
central stars.

We here examined whether the lack of short-period planets
around clump giants could be reproduced by evolutionary effect of
the central stars based on available stellar evolutionary models.
If the clump giants are post-RGB (core helium burning) stars,
short-period planets around them might have been engulfed by the
central stars at the tip of RGB due to tidal torque from the
expanding stellar surface.
We numerically traced tidal evolution of a planetary orbit
($\dot{a}_{tide}$) based on equations from Zahn (1989) and stellar
evolutionary tracks from Lejeune \& Schaerer (2001) (LS01)
and Claret (2004,2006,2007).
We assumed a circular orbit but the result can be applied
to the case of an eccentric orbit by replacing the semimajor axis
with periastron distance.
We also took account of orbital evolution due to mass loss as 
$\dot{a}_{loss}=\dot{M}_{*}a/M_{*}$, where $M_{*}$ is a mass
of the central star. Mass loss of the central star makes planets
move outward because of their weakened gravitational pull on the
planets (e.g., Sackmann et al. 1993; Duncan \& Lissauer 1998).
Thus, net change of orbital radius of a planet is expressed as
$\dot{a}=\dot{a}_{tide}+\dot{a}_{loss}$.
We finally found out, however, that orbital change due to mass
loss is negligible in RGB phase for planets around 2--3 $M_{\odot}$
stars because the mass loss of those stars in RGB phase is negligible
based on the adopted evolutionary tracks.
It excludes the scenario that inner planets were pushed out
to 0.7 AU resulting in the lack of planets within the radius.
The mass loss may be important in the
case of lower-mass stars (Silvotti et al. 2007).
From the orbital calculations, we found out that Jupiter-mass
planets within about 3--4 $R_*$ (radius of a central star) can be
engulfed by the central stars due to the tidal torque during RGB phase.
Our results predict that, around 2--3 $M_{\odot}$ stars, only
planets within 0.2 AU are engulfed by the central stars at the
bottom of RGB ($R_*\lesssim 10R_{\odot}$), but those within
about 0.5 AU can be done at the tip of RGB ($R_*\simeq 25-40
R_{\odot}$). The critical orbital radius at the tip of RGB
for 2$M_{\odot}$ can be larger than 1 AU in the cases of
$Z=0.04$ for LS01 and $Z=0.019$ and 0.04 for Claret's tracks.
When we assume that most of the clump giants are post-RGB stars,
it might be natural that we could not find short-period planets
around them even if they had originally existed.
Typically small orbital eccentricities ($e=0-0.4$) of the planets
discovered around clump giants compared to those around dwarfs
may favor this scenario, in which the planetary orbits could be
tidally circularized during RGB phase.

There can be another possibility, however, that short-period planets
are primordially rare around intermediate-mass stars.
Lack of short-period planets around less evolved subgiants with
1.6--1.9$M_{\odot}$ (Figure \ref{fig-avse}; Johnson et al. 2007b),
which are considered to be first ascent on RGB, may favor this scenario.
Such dependence of orbital distribution of planets on host star's
mass is predicted by Burkert \& Ida (2007). 
They pointed out that in observed data for F--K dwarfs, there may be a
paucity of planets in $a=$0.1--0.6 AU around $\ge1.2M_{\odot}$ stars
(F dwarfs) while the semimajor axis distribution is more uniform
around G and K dwarfs. They showed that the gap could be produced by
shorter viscous diffusion timescale of disks possibly due to smaller
disk size for more massive stars, which can limit the efficiency of 
type II migration of giant planets and keep planets residing close
to initial formation locations beyond snow lines at several AU.
The lack of inner planets around subgiants with 1.6--1.9$M_{\odot}$
might be consistent with this prediction.
It is not clear whether this is also the case for more massive stars
with $\ge 2M_{\odot}$.
Around such stars, due to distant snow line ($>$10 AU) and averagely
large disk mass, the main region for giant planet formation could be
closer to the central stars compared to around low-mass stars.
On the other hand, Kennedy and Kenyon (2007) recently showed that 
the snow line distance changes weakly with stellar mass if they
take account of disk and pre-main-sequence evolution.
More detailed theoretical modeling is required for
planet formation around stars with $\ge 2M_{\odot}$.

In order to further investigate formation and evolution of planetary
systems around massive stars via evolved giants, it is important to
derive their accurate mass and evolutionary status. It is normally
difficult, however, because stars with different mass and evolutionary
status can occupy similar position in the giant branch on the HR diagram.
To overcome this difficulty, asteroseismology will be a powerful tool,
which can probe stellar interior by using tiny stellar oscillations.
Such oscillations were actually detected in some G and K giants
(e.g., Frandsen et al. 2002; Hatzes \& Zechmeister 2007;
Ando et al. 2007).
Applying this technique to the planet-harboring evolved
stars is highly encouraged.
\\

This research is based on data collected at Okayama Astrophysical
Observatory (OAO), which is operated by National Astronomical
Observatory of Japan (NAOJ).
We are grateful to all the staff members of OAO for their support
during the observations. M.I. thanks Dr. A. Claret for his
helpful comments on his evolutionary tracks. 
Data analysis was in part carried out on ``sb'' computer system
operated by the Astronomical Data Analysis Center (ADAC) and
Subaru Telescope of NAOJ. We thank the National Institute
of Information and Communications Technology for their support
on high-speed network connection for data transfer and analysis.
B.S. is supported by Grant-in-Aid for Young Scientists (B) No.17740106,
and H.I., H.A., and M.Y. are supported by Grant-in-Aid for
Scientific Research (C) No.13640247, (B) No.17340056,
(B) No.18340055, respectively,
from the Japan Society for the Promotion of Science (JSPS).
E.T, D.M, and Y.I. are supported by "The 21st Century COE Program: The
Origin and Evolution of Planetary Systems" in Ministry of Education,
Culture, Sports, Science and Technology (MEXT). E.K. and S.I. are
partially supported by MEXT, Japan, the Grant-in-Aid for Scientific
Research on Priority Areas, ``Development of Extra-Solar Planetary
Science.''
This research has made use of the SIMBAD database, operated at
CDS, Strasbourg, France.


\newpage

\onecolumn
\begin{table}[h]
\caption{Stellar parameters}\label{tbl-stars}
\begin{center}
\begin{tabular}{ccccccc}\hline\hline
Parameter      & 18 Del & $\xi$ Aql & HD 81688  & Source/Method \\
\hline			   			   
Sp. Type       & G6 III          & K0 III            & K0 III--IV         & Hipparcos catalogue \\
$\pi$ (mas)    & 13.68$\pm$0.70  & 15.96$\pm$1.01    & 11.33$\pm$0.84     & Hipparcos catalogue \\
$V$            & 5.51            & 4.71              & 5.40               & Hipparcos catalogue \\
$B-V$          & 0.934           & 1.023             & 0.993              & Hipparcos catalogue \\
$A_{V}$        & 0.04            & 0.10              & 0.10               & Arenou et al's (1992) table \\
$M_{V}$        & 1.15            & 0.63              & 0.57               & From $\pi$, $V$, and $A_{V}$ \\
$B.C.$         & $-0.39$         & $-0.48$           & $-0.48$            & Kurucz (1993)'s theoretical calculation \\
$T_{\rm eff}$ (K) & 4979$\pm$18  & 4780$\pm$30       & 4753$\pm$15        & Determined from Fe I and Fe II lines \\ 
$\log g$       & 2.818$\pm$0.060 & 2.66$\pm$0.11     & 2.223$\pm$0.050    & Determined from Fe I and Fe II lines \\
$v_t$          & 1.22$\pm$0.06   & 1.49$\pm$0.09     & 1.43$\pm$0.05      & Determined from Fe I and Fe II lines \\
$[$Fe/H$]$     & $-0.052\pm0.023$ & $-0.205\pm0.039$  & $-0.359\pm$0.020   & Determined from Fe I and Fe II lines \\
$L$ ($L_{\odot}$) & 40           & 69                & 72                 & From $M_{V}$ and B.C. \\
$R$ ($R_{\odot}$) & 8.5          & 12                & 13                 & From $T_{\rm eff}$ and $L$ \\
$M$ ($M_{\odot}$) & 2.3          & 2.2               & 2.1                & Lejeune \& Schaerer's theoretical tracks \\
$v\sin i$ (km s$^{-1}$) & --     & 2.0               & 1.2                & de Medeiros \& Mayor (1999)\\
\hline
\end{tabular}
\end{center}
Note -- While the uncertainties of [Fe/H] are the rms errors
($\equiv\sigma/\sqrt N$) relevant to the average over $N$ lines,
those given for $T_{\rm eff}$, $\log g$, and $v_{\rm t}$, are
nothing but the internal statistical errors (for a given data set
of Fe~{\sc i} and Fe~{\sc ii} line equivalent widths) evaluated
by the procedure described in subsection 5.2 of Takeda et al. (2002).
Actually, since these parameter values are sensitive to slight
changes in the equivalent widths as well as to the adopted set
of lines, realistic ambiguities may be by a factor of $\sim$ 2--3
larger than these estimates from a conservative point of view
(e.g., 50--100 $K$ in $T_{\rm eff}$, 0.1--0.2~dex in $\log g$).
It is normally difficult to precisely determine mass of clump
giants and also set reliable error bars on it because stellar
evolutionary tracks with various mass, metallicity and evolutionary
status occupy similar position near the clump region on the HR
diagram.
Corresponding to the above uncertainties in the parameters of these
stars assumed as $\Delta T_{\rm eff}\sim 100$~K, $\Delta$[Fe/H]
$\sim$ 0.1 dex, and $\Delta L/L_{\odot}\sim$ 5--10\% (mostly
due to Hipparcos parallax errors), errors for the mass and
radius of these stars are estimated to be $\Delta M\sim$
0.2--0.3 $M_{\odot}$ and $\Delta R/R_{\odot}\sim$ 5--10\%.
Moreover, we should keep in mind that the resulting mass value
may appreciably depend on the chosen set of theoretical
evolutionary tracks (e.g., the systematic difference as large
as $\sim 0.5M_{\odot}$ for the case of metal-poor tracks
between Lejeune \& Schaerer (2001) and Girardi et al. (2000).;
cf. footnote 3).
Further comprehensive discussion of stellar parameters of late-G
and early-K giants and their ambiguities, based on a larger
number ($\sim 320$) of sample stars, will be presented in a
forthcoming paper (Takeda et al. in preparation).
\end{table}

\begin{longtable}{ccc}
  \caption{Radial Velocities of 18 Del}\label{tbl-HD199665}
  \hline\hline
  JD & Radial Velocity & Uncertainty\\
  ($-$2450000) & (m s$^{-1}$) & (m s$^{-1}$)\\
  \hline
  \endhead
2489.1422 & 0.8 & 5.5\\
2507.1266 & $-$22.4 & 5.7\\
2541.1260 & $-$56.1 & 5.5\\
2857.1356 & $-$46.5 & 6.1\\
2896.0403 & $-$2.9 & 6.4\\
2927.0518 & $-$9.1 & 5.2\\
2974.9012 & 31.4 & 5.5\\
2994.8978 & 44.1 & 4.5\\
3005.8960 & 33.2 & 8.5\\
3008.8873 & 30.6 & 6.5\\
3077.3450 & 94.8 & 7.5\\
3100.2914 & 86.4 & 6.3\\
3131.3138 & 99.2 & 5.3\\
3201.1162 & 143.8 & 6.5\\
3246.1069 & 129.6 & 5.6\\
3249.1072 & 122.0 & 5.3\\
3284.9244 & 95.7 & 4.3\\
3289.9548 & 95.1 & 4.4\\
3305.9224 & 88.6 & 6.6\\
3310.9473 & 81.8 & 5.1\\
3331.9186 & 87.3 & 5.4\\
3334.8762 & 94.0 & 6.5\\
3340.0094 & 67.5 & 6.0\\
3362.8767 & 81.9 & 4.9\\
3364.9064 & 77.3 & 6.1\\
3428.3705 & 26.5 & 6.3\\
3448.3432 & $-$9.0 & 6.7\\
3474.3188 & $-$51.5 & 6.3\\
3495.2623 & $-$53.3 & 7.6\\
3520.2931 & $-$42.2 & 5.8\\
3527.2996 & $-$43.9 & 9.3\\
3579.1322 & $-$110.6 & 8.8\\
3635.0984 & $-$94.2 & 6.1\\
3655.9467 & $-$123.8 & 4.3\\
3692.9010 & $-$118.8 & 5.6\\
3719.9211 & $-$134.4 & 3.9\\
3726.8798 & $-$132.6 & 5.4\\
3740.8819 & $-$122.1 & 7.0\\
3815.3434 & $-$82.8 & 6.9\\
3833.3331 & $-$90.4 & 6.8\\
3853.2909 & $-$69.7 & 7.4\\
3890.2191 & $-$45.3 & 8.3\\
3938.2715 & 0.1 & 5.7\\
3962.2112 & 29.9 & 6.7\\
4018.0439 & 51.1 & 4.4\\
4048.9950 & 69.9 & 5.9\\
4088.8993 & 87.5 & 3.9\\
4195.3185 & 93.4 & 6.3\\
4216.3152 & 81.6 & 4.8\\
4254.2312 & 100.9 & 5.9\\
4261.2661 & 101.9 & 5.5\\
  \hline
\end{longtable}

\begin{table}
  \caption{Orbital Parameters}\label{tbl-planets}
  \begin{center}
    \begin{tabular}{lrrrrrr}
  \hline\hline
  Parameter & 18 Del & $\xi$ Aql & HD 81688 & HD 104985\\
  \hline
$P$ (days)                    & 993.3$\pm$3.2     & 136.75$\pm$0.25    & 184.02$\pm$0.18     & 199.505$\pm$0.085\\
$K_1$ (m s$^{-1}$)            & 119.4$\pm$1.3     & 65.4$\pm$1.7       & 58.58$\pm$0.97      & 166.8$\pm$1.3\\
$e$                           & 0.08$\pm$0.01     & 0 (fixed)          & 0 (fixed)           & 0.090$\pm$0.009\\
$\omega$ (deg)                & 166.1$\pm$6.5     & 0 (fixed)          & 0 (fixed)           & 203.5$\pm$5.7\\
$T_p$    (JD$-$2,450,000)     & 1672$\pm$18       & 3001.7$\pm$1.4     & 2335.4$\pm$1.1      & 1927.5$\pm$3.3\\
$a_1\sin i$ (10$^{-3}$AU)     & 10.89$\pm$0.11    & 0.824$\pm$0.022    & 0.993$\pm$0.016     & 3.053$\pm$0.024\\
$f_1(m)$ (10$^{-7}M_{\odot}$) & 1.743$\pm$0.054   & 0.0399$\pm$0.0031  & 0.0385$\pm$0.0019   & 0.953$\pm$0.022\\
$m_2\sin i$ ($M_{\rm J}$)     & 10.3              & 2.8                & 2.7                 & 8.3\\
$a$ (AU)                      & 2.6               & 0.68               & 0.81                & 0.95\\
$N_{\rm obs}$                 & 51                & 26                 & 81                  & 52\\
rms (m s$^{-1}$)              & 15.4              & 22.3               & 24.0                & 26.6\\
Reduced $\sqrt{\chi^2}$       & 2.5               & 3.8                & 3.9                 & 4.1\\
  \hline
    \end{tabular}
  \end{center}
\end{table}

\begin{table}
  \caption{Radial Velocities of $\xi$ Aql}\label{tbl-HD188310}
  \begin{center}
    \begin{tabular}{ccc}
  \hline\hline
  JD & Radial Velocity & Uncertainty\\
  ($-$2450000) & (m s$^{-1}$) & (m s$^{-1}$)\\
  \hline
3102.2929 & $-$11.4 & 7.0\\
3213.1702 & $-$85.6 & 7.1\\
3310.9914 & 23.7 & 5.3\\
3448.3230 & $-$50.3 & 4.8\\
3470.3333 & $-$48.7 & 7.2\\
3499.1952 & $-$53.7 & 5.9\\
3520.3046 & 4.5 & 7.9\\
3579.0731 & $-$23.5 & 7.8\\
3607.0846 & $-$46.2 & 8.8\\
3636.0151 & $-$30.2 & 6.5\\
3695.9542 & 80.5 & 6.1\\
3813.3486 & 77.0 & 6.2\\
3887.2842 & $-$32.9 & 5.2\\
3962.0643 & 103.6 & 6.9\\
3967.0716 & 71.7 & 8.7\\
3974.0548 & 59.5 & 6.9\\
3987.0199 & 23.8 & 6.6\\
4018.0165 & $-$45.5 & 6.5\\
4022.0373 & $-$81.2 & 5.2\\
4048.9763 & $-$41.7 & 6.1\\
4088.8857 & 50.7 & 6.1\\
4143.3757 & $-$16.9 & 7.1\\
4172.3496 & $-$57.1 & 5.6\\
4216.3038 & 31.0 & 5.8\\
4220.3112 & 58.2 & 6.9\\
4254.1264 & 8.5 & 5.5\\
  \hline
    \end{tabular}
  \end{center}
\end{table}

\begin{longtable}{ccc}
  \caption{Radial Velocities of HD 81688}\label{tbl-HD81688}
  \hline\hline
  JD & Radial Velocity & Uncertainty\\
  ($-$2450000) & (m s$^{-1}$) & (m s$^{-1}$)\\
  \hline
  \endhead
2311.1798 & 50.7 & 4.4\\
2335.2204 & 44.8 & 7.1\\
2361.2050 & 7.7 & 7.4\\
2656.0534 & 32.3 & 6.5\\
2677.2518 & 50.7 & 5.1\\
2709.1489 & 61.3 & 4.9\\
2923.3218 & 38.6 & 5.5\\
2974.2203 & $-$40.3 & 5.2\\
2991.2026 & $-$50.4 & 5.8\\
3002.1804 & $-$56.3 & 5.7\\
3006.1215 & $-$67.4 & 6.1\\
3024.0642 & $-$36.3 & 6.2\\
3028.0500 & $-$36.5 & 6.5\\
3052.0118 & $-$2.5 & 5.9\\
3056.1429 & $-$14.7 & 6.6\\
3075.0432 & 68.7 & 4.9\\
3078.2002 & 92.9 & 4.7\\
3100.0052 & 43.4 & 6.5\\
3104.1309 & 15.4 & 6.8\\
3105.1169 & 43.8 & 7.1\\
3106.0327 & 0.2 & 6.9\\
3107.1472 & $-$8.8 & 6.1\\
3110.1677 & $-$11.2 & 7.1\\
3112.0892 & $-$40.5 & 6.7\\
3123.0991 & $-$23.6 & 9.9\\
3131.0036 & $-$13.5 & 6.5\\
3131.9477 & $-$13.7 & 6.1\\
3136.0225 & $-$19.2 & 6.9\\
3154.0323 & $-$66.1 & 7.9\\
3157.9805 & $-$69.5 & 6.7\\
3159.9783 & $-$69.4 & 6.8\\
3162.0172 & $-$97.9 & 7.7\\
3284.3176 & 33.5 & 5.5\\
3307.2533 & 34.1 & 6.8\\
3311.2119 & $-$7.4 & 5.6\\
3332.3369 & $-$90.5 & 4.9\\
3333.2075 & $-$41.4 & 5.2\\
3334.2916 & $-$81.4 & 5.0\\
3335.2195 & $-$28.0 & 5.3\\
3336.2558 & $-$48.1 & 4.8\\
3338.3409 & $-$58.4 & 4.7\\
3339.3451 & $-$42.1 & 5.2\\
3340.2926 & $-$46.9 & 4.9\\
3361.3194 & $-$45.9 & 6.1\\
3362.3774 & $-$65.0 & 7.6\\
3363.3370 & $-$46.3 & 6.1\\
3364.2794 & $-$33.6 & 6.0\\
3365.2714 & $-$21.5 & 6.0\\
3366.1669 & $-$40.4 & 6.4\\
3367.1279 & $-$26.2 & 6.1\\
3378.3533 & $-$37.7 & 5.1\\
3402.2015 & 13.0 & 5.3\\
3407.2808 & 43.0 & 4.4\\
3424.0331 & 22.4 & 6.3\\
3445.1913 & 40.4 & 6.8\\
3449.0443 & 64.7 & 5.1\\
3468.1174 & 22.1 & 9.0\\
3495.0446 & $-$47.4 & 7.0\\
3659.3268 & 32.7 & 5.5\\
3694.3233 & $-$25.8 & 4.9\\
3720.3760 & $-$55.6 & 6.1\\
3728.3348 & $-$60.7 & 6.0\\
3743.2486 & $-$47.5 & 5.4\\
3775.1171 & 73.4 & 4.0\\
3811.2229 & 41.7 & 5.4\\
3812.0871 & 44.9 & 4.7\\
3831.1137 & 70.8 & 9.3\\
3853.1140 & $-$21.3 & 6.8\\
3886.9814 & $-$101.0 & 9.0\\
3888.9868 & $-$97.2 & 8.4\\
4018.3495 & 54.9 & 4.4\\
4049.2922 & $-$9.6 & 5.7\\
4087.3587 & $-$62.3 & 5.7\\
4092.2203 & $-$35.6 & 5.7\\
4115.3330 & $-$9.9 & 6.0\\
4122.2220 & $-$21.2 & 5.7\\
4127.1958 & $-$37.0 & 8.0\\
4143.2036 & 23.2 & 4.5\\
4195.1511 & 40.0 & 6.3\\
4216.0967 & 16.8 & 6.7\\
4255.9738 & $-$59.0 & 6.3\\
  \hline
\end{longtable}

\begin{longtable}{ccc}
  \caption{Radial Velocities of HD 104985}\label{tbl-HD104985}
  \hline\hline
  JD & Radial Velocity & Uncertainty\\
  ($-$2450000) & (m s$^{-1}$) & (m s$^{-1}$)\\
  \hline
  \endhead
2284.3135 & $-$97.5 & 6.6\\
2337.2114 & $-$110.8 & 6.8\\
2425.9960 & 153.5 & 7.7\\
2426.9682 & 159.7 & 5.7\\
2427.9753 & 117.4 & 7.7\\
2484.9871 & $-$129.7 & 8.7\\
2505.9601 & $-$150.1 & 5.8\\
2592.2476 & 174.6 & 9.8\\
2592.3501 & 173.1 & 7.6\\
2638.3674 & 94.0 & 7.1\\
2651.3383 & 59.4 & 7.4\\
2677.1079 & $-$63.0 & 7.9\\
2680.2379 & $-$49.3 & 8.0\\
2692.2555 & $-$163.8 & 7.1\\
2706.1590 & $-$198.4 & 8.6\\
2710.1659 & $-$153.1 & 6.3\\
2735.1287 & $-$120.6 & 7.8\\
2756.1053 & $-$32.4 & 9.5\\
2974.2742 & 74.2 & 7.5\\
3002.2274 & 180.6 & 6.1\\
3024.1295 & 155.2 & 5.7\\
3030.3033 & 110.8 & 5.6\\
3052.0835 & 26.3 & 7.0\\
3077.0500 & $-$32.3 & 5.2\\
3100.0693 & $-$135.6 & 5.2\\
3131.0540 & $-$154.7 & 5.7\\
3160.9710 & 70.7 & 5.7\\
3332.3794 & $-$155.5 & 6.2\\
3335.3282 & $-$130.7 & 6.7\\
3340.3146 & $-$129.6 & 6.2\\
3363.3593 & 44.4 & 6.0\\
3367.2023 & 53.9 & 5.9\\
3403.3051 & 174.8 & 6.3\\
3447.2055 & 68.0 & 6.3\\
3694.3459 & $-$175.4 & 5.6\\
3729.3337 & $-$133.6 & 6.3\\
3745.3168 & $-$33.4 & 6.8\\
1982.1212 & 58.4 & 6.2\\
1982.1341 & 77.4 & 4.9\\
3775.2303 & 16.7 & 7.1\\
3812.1245 & 139.3 & 4.3\\
3889.0249 & $-$161.0 & 7.0\\
2016.0945 & 157.3 & 5.4\\
2033.0394 & 122.9 & 6.8\\
2041.0056 & 123.2 & 6.8\\
2042.0112 & 95.1 & 5.0\\
4049.3406 & 64.8 & 6.4\\
4088.3821 & $-$111.6 & 5.8\\
4126.3022 & $-$186.5 & 5.4\\
4143.2217 & $-$63.1 & 6.9\\
4169.0809 & 46.6 & 5.6\\
4216.1184 & 131.1 & 7.1\\
  \hline
\end{longtable}

\begin{table}
  \caption{Bisector Quantities}\label{tbl-bisector}
  \begin{center}
    \begin{tabular}{lrrrrrr}
  \hline\hline
  Bisector Quantities & 18 Del & $\xi$ Aql & HD 81688 & HD 104985\\
  \hline
Bisector Velocity Span (BVS) (m s$^{-1}$) & 0.5$\pm$3.9     & $-$3.9$\pm$5.2    & 6.4$\pm$4.8     & 10.7$\pm$4.9\\
Bisector Velocity Curve (BVC) (m s$^{-1}$) & 0.0$\pm$1.7     & $-$1.2$\pm$1.8       & 4.0$\pm$3.3      & $-$3.6$\pm$3.2\\
Bisector Velocity Displacement (BVD) (m s$^{-1}$) & $-$205.8$\pm$10.1  & $-$127.3$\pm$12.4  & $-$139.37$\pm$14.3 & $-$323.7$\pm$16.5\\
  \hline
    \end{tabular}
  \end{center}
\end{table}

\begin{table}
  \caption{Substellar Companions around Evolved Stars}\label{tbl-pllist}
  \begin{center}
    \begin{tabular}{llrrrrrrrr}
  \hline\hline
  HD &  & Sp. Type &     $M_{*}$ ($M_{\odot}$) & $R_{*}$ ($R_{\odot}$) & $M_{p}\sin i$ ($M_J$) & $a$ (AU) & $e$ & Ref.\\
  \hline
Clump Giants\\
        & NGC 4349       &           & 3.9   & --    & 19.8   & 2.38     & 0.19    & 1\\
13189   &                & K2 II     & 2--7  & --    &  8--20 & 1.5--2.2 & 0.28    & 2\\
28305   & $\epsilon$ Tau & K0 III    & 2.7   & 13.7  & 7.6    & 1.93     & 0.15    & 3\\
107383  & 11 Com         & G8 III    & 2.7   & 19    & 19.4   & 1.29     & 0.23    & 4\\
        & NGC 2423       &           & 2.4   & --    & 10.6   & 2.1      & 0.21    & 1\\
199665  & 18 Del         & G6 III    & 2.3   & 8.5   & 10.3   & 2.6      & 0.08    & 5\\
104985  &                & G9 III    & 2.3   & 11    &  8.3   & 0.95     & 0.09    & 5,6\\
17092   &                & K0        & 2.3   & 10.1  &  4.6   & 1.29     & 0.166   & 7\\
188310  & $\xi$ Aql      & K0 III    & 2.2   & 12    &  2.8   & 0.68     & 0       & 5\\
81688   &                & K0 III--IV& 2.1   & 13    &  2.7   & 0.81     & 0       & 5\\
11977   &                & G5 III    & 1.91  & 10.1  &  6.54  & 1.93     & 0.4     & 8\\
62509   & $\beta$ Gem    & K0 III    & 1.7   & 8.8   &  2.3   & 1.6      & 0.02    & 9\\
\hline
Subgiants\\
210702  &                & K1 IV     & 1.85  & 4.72  &  2.0   & 1.17     & 0.152   & 10\\
192699  &                & G8 IV     & 1.68  & 4.25  &  2.5   & 1.16     & 0.149   & 10\\
175541  &                & G8 IV     & 1.65  & 3.85  &  0.61  & 1.03     & 0.33    & 10\\
\hline
Low-mass K Giants\\
222404  & $\gamma$ Cep   & K0 III    & 1.59  & 4.66  &  1.7   & 2.13     & 0.12    & 11\\
122430  &                & K3 III    & 1.39  & 22.9  &  3.71  & 1.02     & 0.68    & 12\\
73108   & 4 UMa          & K1 III    & 1.23  & 18.1  &  7.1   & 0.87     & 0.432   & 13\\
47536   &                & K1 III    & 1.1   & 23.47 &  4.96  & 1.61     & 0.2     & 14\\
137759  & $\iota$ Dra    & K2 III    & 1.05  & 12.9  &  8.9   & 1.3      & 0.7     & 15\\
  \hline
    \end{tabular}
  \end{center}
References.-- (1) Lovis \& Mayor (2007); (2) Hatzes et al. (2005); (3) Sato et al. (2007);
(4) Liu et al. (2007); (5) This work; (6) Sato et al. (2003); (7) Niedzielski et al. (2007);
(8) Setiawan et al. (2005); (9) Hatzes et al. (2006); (10) Johnson et al. (2007);
(11) Hatzes et al. (2003); (12) Setiawan (2003); (13) D$\ddot{\rm{o}}$llinger et al. (2007);
(14) Setiawan et al. (2003); (15) Frink et al. (2002)
\end{table}

\end{document}